\documentclass[12pt,twocolumn]{article}
\usepackage{flushend}
\usepackage{cuted}
\usepackage{color}
\usepackage{graphicx}
\usepackage{hyperref}
\usepackage{pstricks,amssymb}
\usepackage{fancyhdr}
\usepackage[raggedright]{titlesec}
\begin{document}
\title{The genesis of the internal resistance of a battery -- a physical perspective
\\
\vspace{.3cm}
\large{Ashok K. Singal}\\
\vspace{0.3cm}
\normalsize{
Astronomy and Astrophysics Division\\ 
Physical Research Laboratory\\
Navrangpura, Ahmedabad - 380 009, India.\\
{\small asingal@prl.res.in}\\
\vspace{0.3cm}
(Submitted 14-11-2015)}\\
\rule[0.1cm]{16cm}{0.02cm} \\
\textbf{Abstract}\\
\flushleft \normalsize {The standard exposition of the internal resistance of a battery, 
that a battery comprises a source of emf in series with an internal resistance, as given 
in engineering and physics text-books, is lacking in proper explanation. It is treated merely as 
an experimental fact, and not something that should follow from logic. The battery has a tendency 
to maintain electric potential difference across its terminals equal to its chemical potential, 
and in an open circuit, when no current flows, these two do match. However in a closed circuit, 
a drop in electric potential across the battery terminals is inevitable for a steady flow of electric 
current throughout the circuit, because the chemical reactions driving the electric current within 
the battery can proceed only if the electric potential at its terminals differs from the chemical potential. 
It is shown that for small voltage changes, the current passing through the battery is linearly proportional 
to the change in potential from the open-circuit value (i.e., its chemical potential), giving rise to a 
semblance of an internal resistance in series with the external resistance. It follows that a battery {\em has 
to have} an internal resistance in order to function as a power source. It is also shown that Thevenin's theorem 
does not make our results superfluous, in fact our results are presupposed in its derivation.
} \\
%\rule[0.1cm]{16cm}{0.02cm} 
}
\date{}
\maketitle
\thispagestyle{fancy} 
\lhead{\textbf{Physics Education}}
\chead{\thepage}
\rhead{\bf {dateline}(to be added by Editor)}
\lfoot{Volume xx, Number y Article Number : n.(to be added by editor) }
\cfoot{ }
\rfoot{www.physedu.in}
\renewcommand{\headrulewidth}{0.4pt}
\renewcommand{\footrulewidth}{0.4pt}
%%
%--------------------------------------------------
\section{Introduction}
In almost all physics or engineering undergraduate text-books \cite{1,3,4,5}, the internal resistance of a battery is introduced 
more or less as a factual statement that a battery comprises a source of emf in series with an internal resistance $R_i$ (Fig.~(1a)), 
which is the resistance of the electrolyte of the battery. In general there is not much exposition as to the genesis of the internal 
resistance of the battery and more specifically why it needs to be put outside the battery in series with the external resistance. 
A student soon learns to live with it and, at most taking it as an experimental fact, moves on. But a feeling remains that something 
is lacking. After all the internal resistance is due to the constituents of the battery within it, therefore the word ``internal resistance'' 
conjures up a vision like that of a resistance internal to the battery like in Fig.~(1b), or equivalently where the emf of the battery is across 
its internal resistance. Then with a finite voltage at its ends (positive and negative electrodes of the battery) one expects from Ohm's 
law that there should be a finite current flowing through the resistor within the battery (even if the external circuit were open) as long 
as a finite electric potential difference exists across the resistor. Further, in an actual circuit when the circuit is closed and an electric 
current does flow through the internal resistance, it is in a direction from a lower electric potential (negative electrode) to a higher one 
(positive electrode) within the battery (Fig.~(1c)), contrary to that expected from Ohm's law for a normal resistor where the electric current 
should flow from higher to a lower electric potential which is not seen within the battery. Additionally, as the internal resistance is 
supposedly that of the electrolyte residing in-between the two electrodes of the battery, how come the internal resistance is shown 
to exist not between the two electrodes as in Fig.~(1b), but -- somewhat mysteriously -- is instead put in series with the external resistance 
outside the battery Fig.~(1c)?  Moreover, why should it be causing a drop in potential from the open-circuit value when the internal resistance 
itself is a part and parcel of the battery system, giving rise to that potential? Here we should clarify that we are not doubting the truth of the 
long-known experimental facts (for example, we know experimentally that Fig.~(1a) and Fig.~(1c) are factually correct while Fig.~(1b) is not), we are 
only attempting to logically examine these facts from a simple physical perspective.
\begin{figure*}
\scalebox{0.9}{\includegraphics{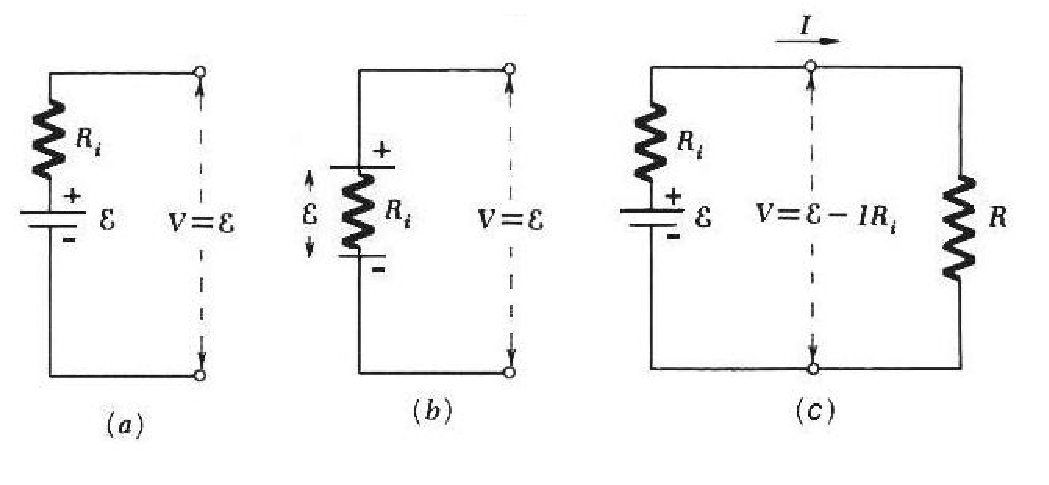}}
\caption{The voltages and currents in a battery of emf ${\cal E}$ for (a) an open circuit with internal resistance 
$R_i$ in series outside the battery (b) $R_i$ ``inside'' the battery and (c) a closed circuit with $R_i$ in series 
with the external resistance.}
\end{figure*}
\section{EMF, voltage and current}
A battery or cell is an electrochemical device \cite{6} that converts chemical energy to electric 
energy by driving an electric current through a circuit containing a load (an external resistance).
Historically, a single chemical 
source of emf was called a cell and a set of interconnected cells was called a battery, 
however, it is now common practice to refer to even a single cell as a battery, 
as is done here too. Batteries could be of different kinds, for example, disposable ones designed to be used only once 
and the rechargeable ones, which can be used more than once. 
A common example of the former is the zinc-carbon battery, often used in torch-lights. 
The rechargeable batteries include lead-acid batteries used in automobiles while others like 
nickel-cadmium or lithium-ion are used in mobile phones and laptop computers. A battery in general, 
consists of two electrodes of different material immersed in an  electrolyte, which could be 
a fluid or a moist paste. The electrolyte interacts chemically with the electrodes and due to their 
chemical reactions a push is exerted on the positive charges towards one terminal, called the positive 
electrode, and on the negative charges towards the other terminal, called the negative electrode. 
Irrespective of the make of a battery (its type, size, volume, the nature of the electrodes and the
electrolyte and the details of their chemical reactions etc.), a battery ultimately causes a separation of 
positive and negative charges, giving rise to an electric potential across the battery. 
The chemical potential, which is the line integral of the force per unit charge due to chemical reactions 
(from the negative electrode to the positive electrode), is called the emf ${\cal E}$ of the battery
(historically called electromotive force which actually is a misnomer as ${\cal E}$ is not a force but is 
instead a potential difference between the two electrodes).  To a first approximation we can write the effective 
force due to chemical reactions on a charge $e$ as $F_{c}=e{\cal E}/d$, where $d$ is the  
distance between the two electrodes. It is this force $F_c$ due to the chemical reactions 
that pushes positive charges towards the positive electrode and the negative charges towards the negative 
electrode inside the battery or cell, giving rise to an electric potential difference between the two electrodes. 
In general the change in electric potential within the battery may not be linear and is mostly 
localized at the electrode--electrolyte interfaces, 
but from energetic point of view what finally matters is the net potential difference $V$ between the two 
electrodes. This in turn gives rise to an electric field $E=-V/d$ within the battery which exerts on every charge $e$ an 
electric force $eE$ in a direction opposite to the force $F_c$ due to the chemical reactions. As a result the net 
force on a charge pushing it towards its respective electrode within the battery becomes,
\begin{equation}
F = F_c+ eE =e \frac {{\cal E}-V}{d}.
\end{equation}
As long as $V$ is smaller than ${\cal E}$, the force $F>0$ and the charges will continue to move inside the battery 
toward their respective electrodes, with more and more charges getting deposited there. However with increasing $V$, 
$F$ will decrease, reducing the current flow inside the battery, until the electrodes achieves a voltage difference 
$V={\cal E}$. Then from Eq. (1), $F=0$ and the charge movement reduces to zero inside the battery. Thus in an open circuit 
the battery will acquire across its terminals a voltage $V$ equal to its emf ${\cal E}$, i.e. its chemical potential 
(Fig.~(1a)), with no net force on the charges and hence no electric current within the battery in spite of the electric 
potential $V$ across its terminals. 

Now let a load (an external resistor $R$) be connected across the battery, closing the circuit. Immediately an 
electric current will start from the positive terminal towards the negative one through the external resistance.  
Actually the electric current in a circuit is due to the flow of electrons in a direction opposite to that 
of the current conventionally shown in a circuit (Fig. (1c)), but it does not alter the physics of the problem. 
The external current causes a deficiency of some negative charges at the negative electrode as well as 
neutralizes some positive 
charges at the positive electrode, the reduction of charges causing a slight drop in voltage from the initial 
open-circuit value $V={\cal E}$. This means that the electric field within the battery will now be less than the 
open-circuit value (i.e., $V<{\cal E}$ or from Eq. (1), $F=F_c+eE>0$) and the electric force will not fully cancel the force $F_c$ on the 
charges within the battery due to the chemical reactions. This in turn will cause the charges to move according to 
Eq.~(1) giving rise to a positive current from the negative electrode to the positive electrode inside the battery. 
Initially as the current within the battery may be less than that in the 
external circuit, the charges getting replenished at the terminals will be less in quantity than those 
getting depleted by the external current flow, therefore the voltage $V$ will be falling still further. 
And as ${\cal E}-V$ increases, this should give rise to not only a still higher current within the battery, 
it will also cause a drop, even if only slight, in the external current as the voltage $V$ across the 
external resistor drops. A stage however, will be reached very soon when the internal 
current within the battery becomes equal to that in the external circuit. Now onwards there will be no 
further change in the voltage at the battery terminals and a state of equilibrium has been reached.
However it will remain a constant struggle for the battery, through the internal current, to keep replenishing 
charges being depleted at its terminals by the external current. 
Thus we see that a current will be flowing from a lower electric potential to the higher one 
within the battery because of the larger push on the charges in that direction by the force $F_c=e{\cal E}/d$ 
due to the chemical reactions than the opposing force by the electric field $eV/d$ from a higher to a lower potential. 
\begin{figure}
\scalebox{0.9}{\includegraphics{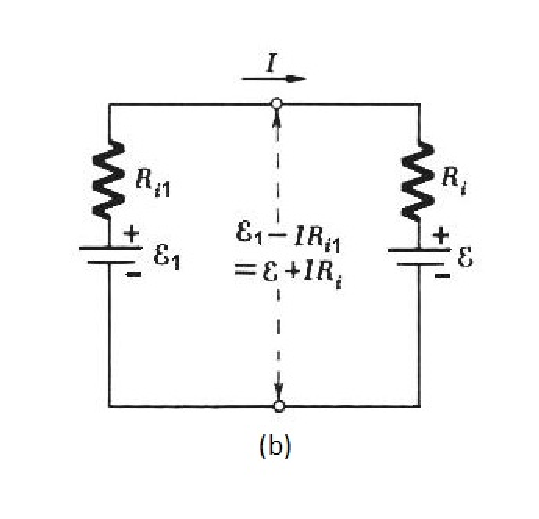}}
\caption{The voltages and currents in a battery of emf ${\cal E}$ when recharging with another 
battery of emf ${\cal E}_1 >{\cal E}$.}
\end{figure}
\section{The internal resistance}
%-----------------
In a closed circuit the electric current flowing within the battery is due to the chemical reactions, 
which will take place only if the voltage across the battery $V$ is different from the chemical potential ${\cal E}$. 
A steady state means the internal current within the battery must be equal to the current $I=V/R$ through the 
external resistance $R$. The internal resistance could be defined by 
%$1/R_i=-({\rm d} I/{\rm d} V)_{V = {\cal E}}$, the negative sign indicating that the current increases when $V$ decreases. 
$R_i=1/\mid{\rm d} I/{\rm d} V\mid_{V = {\cal E}}$. It should be noted that the current increases when $V$ decreases. 
Assuming a constant ${\rm d} I/{\rm d} V$ in a certain range of $V$ around ${\cal E}$ 
and noting that $I=0$ when $V={\cal E}$, we get $R_i=({\cal E}-V)/I$. 
From this we could write $V={\cal E}-I R_i$, which justifies representing the battery as a source of emf ${\cal E}$ 
with its internal resistance $R_i$ in series (Fig.~(1c)). The current $I=V/R$ is then given by $I={\cal E}/(R+R_i)$.
One could even have a reverse current through the battery when $V$ across the battery is made higher 
than ${\cal E}$. For that 
another source of emf, say ${\cal E}_1$ and with an internal resistance $R_{i1}$, so that ${\cal E}_1>{\cal E}$ 
of the battery in question, is connected across its terminals (Fig.~(2)). This is done, for example, to recharge the
lead-acid battery or other rechargeable batteries.  The magnitude of the reverse current through the battery will now be 
given by $I=(V-{\cal E})/R_i$, as the recharging voltage $V={\cal E}_1-I R_{i1}$ is larger than ${\cal E}$, then 
$I=({\cal E}_1-{\cal E})/(R_i+R)$.
The reverse current means that the positive charges move towards the negative electrode while the negative charges 
move towards the positive electrode, thereby reversing the chemical reaction and recharging the battery. 
In the case of non-rechargeable battery no reverse current takes place and we could say it has a discontinuity 
in its internal resistance at $V={\cal E}_{+0}$. (One is in general, 
cautioned against attempts to recharge non-rechargeable batteries as these could explode).

Now $e{\cal E}$ is the amount of chemical energy 
expended as work on a charge $e$ in transporting it from one electrode to the other. Out of this, an amount $eV$ is
spent against the electric field, which ultimately gets delivered to the external load, 
the remaining energy $e ({\cal E}-V)$ represents the ohmic losses within the battery. Thus $({\cal E}-V) I=I^2R_i$ are 
the power losses in the battery as expected from a resistance $R_i$ lying outside the battery in series.

The actual value of the internal resistance of a cell may depend upon a combination of various factors. If the effective
cross-section areas of the electrodes are large, more current may flow through the battery even for the same  
${\cal E}-V$ change, implying a lower $R_i$ value. Similarly a larger separation between the electrodes would imply 
a smaller push on the charges  even for the same ${\cal E}-V$ change (Eq. 1), resulting in a smaller current, 
implying a higher $R_i$ value. The nature of the constituents (electrodes and the electrolyte) of a battery also 
matter as a better conducting electrolyte means a higher current for the same ${\cal E}-V$ and thereby a smaller $R_i$.
Further as with usage the density of chemical components within the battery may decrease, it would lead to an increase 
in the internal resistance. 
%--------------------------------
\section{Thevenin's theorem}
Thevenin's theorem \cite{2} states that any two-terminal network containing energy sources (generators) 
and impedances can be replaced with an equivalent circuit consisting of a voltage source in series with an impedance.
Thus at a first look it may appear to preempt all our above discussion, which may in fact appear redundant. 
But a careful look at the proof of the Thevenin's theorem shows that our above results are rather presupposed there. 
In the proof offered \cite{2} one may have batteries/generators and impedances in series or in 
parallels or in other complicated distributions but to begin with one always has the {\em internal impedances}, if any, 
of the batteries/generators always in series with them. Therefore Thevenin's theorem does not make our results 
superflous, in fact our results are made use of in its derivation. 
\section{Conclusions}
We have shown that due to the tendency of the battery to attain a voltage across its terminals equal to the 
chemical potential ${\cal E}$, a finite drop in voltage from the open circuit value ${\cal E}$ is essential for a steady current in 
the circuit because then and only then will there be chemical reactions taking place so that a current flows 
within the battery. Thus a drop in the voltage is essential for a steady state current implying the existence 
of a finite internal resistance in any practical battery, which can be justifiably represented as a source of 
emf  ${\cal E}$ with a resistance $R_i$ in series. Therefore a battery has to have an internal resistance in order to function 
as a power source. Further we have shown that Thevenin's theorem does not make our results superfluous, 
in fact our results are made use of in its derivation. 
\section*{Acknowledgements}
It was due to the nagging doubts expressed by Tanmay Singal, then an undergraduate student, 
in the standard text-book explanation for the internal resistance of a battery that prompted me to look 
for a more compelling physically arguments. It is hoped these would be beneficial to others as well in 
similar predicaments.
%\section*{References}

\end{document}